%% file: eprint.tex
\documentclass[12pt]{article}
\usepackage{graphicx}

\newcommand\pubnumber{}
\newcommand\pubdate{}

\def\napoli{Department of Physics and Astronomy, University of California Los Angeles, \\ 475 Portola Plaza, Los Angeles, CA 90095, USA}

\def\Title#1{\begin{center} {\Large #1 } \end{center}}
\def\Author#1{\begin{center}{ \sc #1} \end{center}}
\def\Address#1{\begin{center}{ \it #1} \end{center}}

\newcommand\pubblock{\rightline{\begin{tabular}{l} \pubnumber\\
         \pubdate  \end{tabular}}}
\newenvironment{Abstract}{\begin{quotation}  }{\end{quotation}}
\newenvironment{Presented}{\begin{quotation} \begin{center} 
             PRESENTED AT\end{center}\bigskip 
      \begin{center}\begin{large}}{\end{large}\end{center} \end{quotation}}

\input econfmacros.tex

\begin{document}
\begin{titlepage}
\pubblock

\vfill
\Title{Searching for the QCD  Critical Point through Fluctuations \\ at RHIC}
\vfill
\Author{ Roli Esha}
\Address{\napoli}
\vfill
\begin{Abstract}
Fluctuations and correlations of conserved quantities (baryon number, strangeness, and charge) can be used to probe phases of strongly interacting QCD matter and the possible existence of a critical point in the phase diagram. The cumulants of the multiplicity distributions related to these conserved quantities are expected to be sensitive to possible increased fluctuations near a critical point and ratios of the cumulants can be directly compared to the ratios of the susceptibilities from Lattice QCD calculations. In these proceedings, the measurements of the cumulants of net-proton multiplicity distributions from Au+Au collisions at $\sqrt{s_{NN}}$ = 7.7, 11.5, 14.5, 19.6, 27, 39, 62.4 (up to fourth order) and 200 GeV (up to sixth order) as measured by the STAR experiment at RHIC will be presented. Multi-particle correlation functions will also be presented.

The measurement of higher-order cumulants are sensitive to experimental artifacts. Current efficiency correction methods are based on the assumption that the tracking efficiency is strictly binomial. An unfolding technique is used to account for multiplicity dependent detector responses and efficiency variations. A large sample of AMPT events will be used to check the validity of the unfolding technique. The comparison of the various correction approaches should provide important guidance towards a reliable experimental determination of the multiplicity cumulants.
\end{Abstract}
\vfill
\begin{Presented}
Thirteenth Conference on the Intersections of Particle and Nuclear Physics\\
Palm Springs, CA, USA,  May 28--June 3, 2018
\end{Presented}
\vfill
\end{titlepage}
\def\thefootnote{\fnsymbol{footnote}}
\setcounter{footnote}{0}

\section{Introduction}
The main goal of high-energy heavy-ion collisions has been to investigate QCD at high temperature and baryon density. At ordinary temperatures, the quarks and gluons are confined within hadrons, but at very high temperatures and densities, we have a deconfined phase of quarks and gluons, the Quark Gluon Plasma (QGP). Over the past years, evidence for the distinct phases of QGP and hadron gas has been established experimentally. \\

The Beam Energy Scan program at RHIC allows us to study the QCD phase diagram by varying the collision energy for heavy ions, hence, scanning baryon chemical potential ($\mu_B$) and temperature ($T$). For vanishing baryon chemical potential, the transition from the QGP phase to the Hadron Gas phase is a smooth crossover~\cite{lat1}, while QCD-inspired models predict an existence of first-order phase transition for large baryon chemical potential. Thermodynamic principles, hence, suggest the existence of a critical point in the QCD phase diagram. \\

Theoretically, event-by-event fluctuations of conserved quantities, namely charge, baryon number and strangeness, is used to probe the critical phenomenon in the QCD phase diagram. Experimentally, this translates to the measurement of the cumulants of event-by-event net-particle multiplicity distributions - net-charge, net-proton (proxy for net-baryon) and net-kaon (proxy for net-strangeness). Correlation function of various particles can also be obtained from the same. The cumulants ($C_n$) are related to the susceptibility ($\chi^{(n)}$) of the system~\cite{int1}, which is the derivative of free energy with respect to the chemical potential, in the following way
\begin{equation}
\chi_q ^{(n)} = \frac{\partial^n (p/T^4)}{\partial (\mu_q/T)^n} = \frac{1}{VT^3} \times C_{n,q},
\end{equation}
where $V$ is the volume, $T$ is the temperature, $p$ is the pressure, and $q$ denotes the conserved quantity, that is, charge, baryon number or strangeness. The ratios of such cumulants as experimental observables cancel the volume and temperature dependence and can be directly compared to the ratios of susceptibilities from theoretical calculations. The higher-order cumulants were predicted to be more sensitive to the signatures of phase transition~\cite{int2}, and perhaps more susceptible to experimental artifacts as well. 

In these proceedings, the cumulants for net-proton and net-kaon multiplicity will be presented from the data taken in the years 2010 and 2011 by the STAR experiment at the Relativistic Heavy Ion Collider (RHIC) at the Brookhaven National Laboratory. The measurement of cumulants of net-charge distribution will be presented by both the STAR and the PHENIX experiments.

\section{Analysis details}
The net-particle distributions are constructed from the event-by-event difference of positively and negatively charged particles. The analysis is carried out using minimum bias events obtained after rejecting pile-up and other background interactions. All the tracks in an event that are included in the analysis also undergo some quality checks.

In order to get the precise measurements of the event-by-event fluctuations of conserved quantities and suppress background, a series of analysis techniques are applied~\cite{tec1}. These can be summarized as the following:
\begin{enumerate}
\item The centrality of the collision is determined by excluding the particle of interest. This helps avoid auto-correlation, which is a background effect and can reduce the magnitude of the signal in fluctuation measurements~\cite{tec2}.
\item Centrality Bin Width Correction is applied to suppress volume fluctuation with a wide centrality range, which could lead to artificial centrality dependence for the observables. For this, the cumulants are obtained as the weighted average for the given centrality~\cite{tec3}.
\item Statistical error is estimated by using either the Bootstrap method~\cite{tec4}, which is based on resampling, or the Delta theorem~\cite{tec5}, which is an analytic technique.
\item As no detector is perfect, an important step in the measurement of cumulants is efficiency correction. The principal idea is to express them in terms of the factorial moments~\cite{tec6} or factorial cumulants~\cite{tec7}, which can be efficiency corrected assuming a Binomial response function. 
\end{enumerate}

\begin{figure}[htb]
\centering
\includegraphics[height=2.0in]{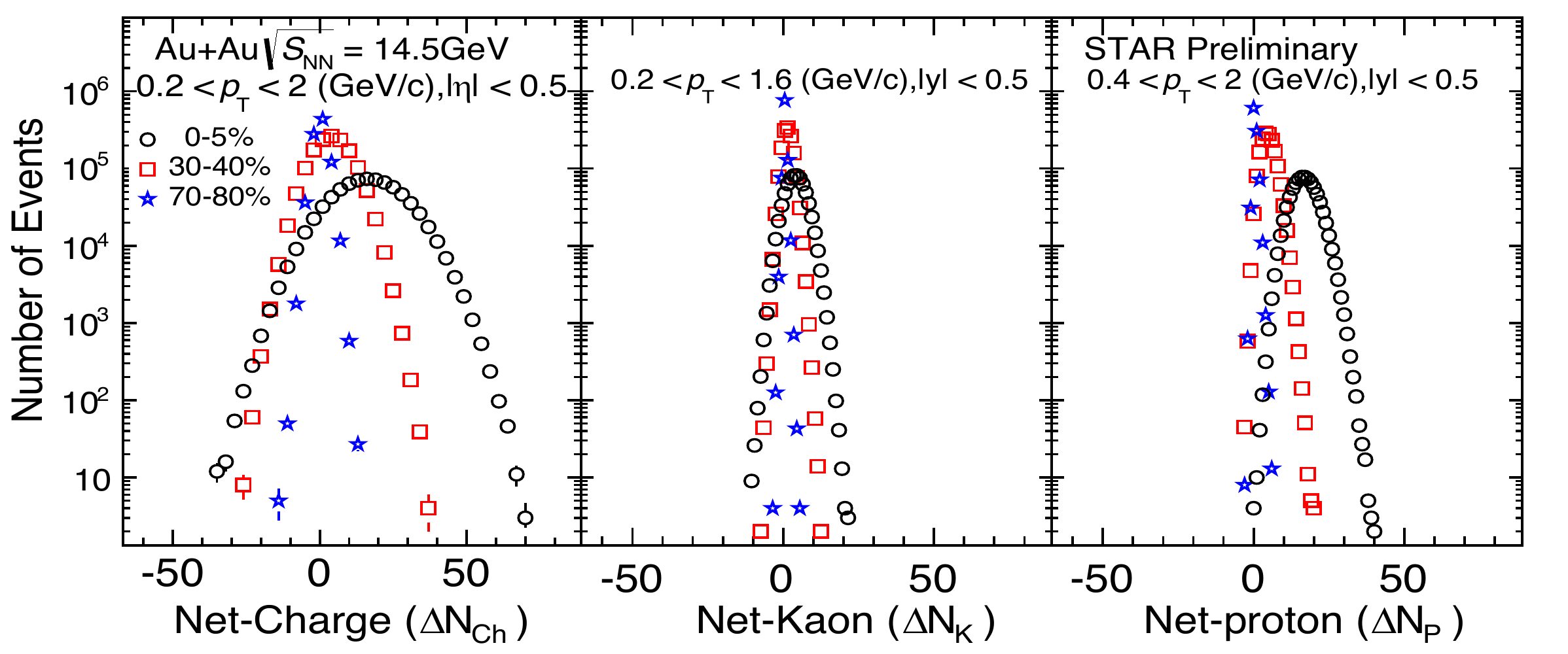}
\caption{Uncorrected raw event-by-event net-particle multiplicity distribution for Au+Au collisions at $\sqrt{s_{NN}} = 14.5$ GeV}
\label{fig:raw}
\end{figure}

For the purpose of illustration, Figure~\ref{fig:raw} shows the raw, uncorrected distribution of the net-charge, net-kaon and net-proton multiplicity distributions for Au+Au collision at $\sqrt{s_{NN}} = 14.5$ GeV for different centralities as measured by the STAR experiment~\cite{fig1}. The black circles are for 0-5\% central collisions, the red squares for 30-40\% and blue stars for 70-80\% central collisions.

\section{Results and discussions}
\begin{figure}[htb]
\centering
\includegraphics[width=1.96in]{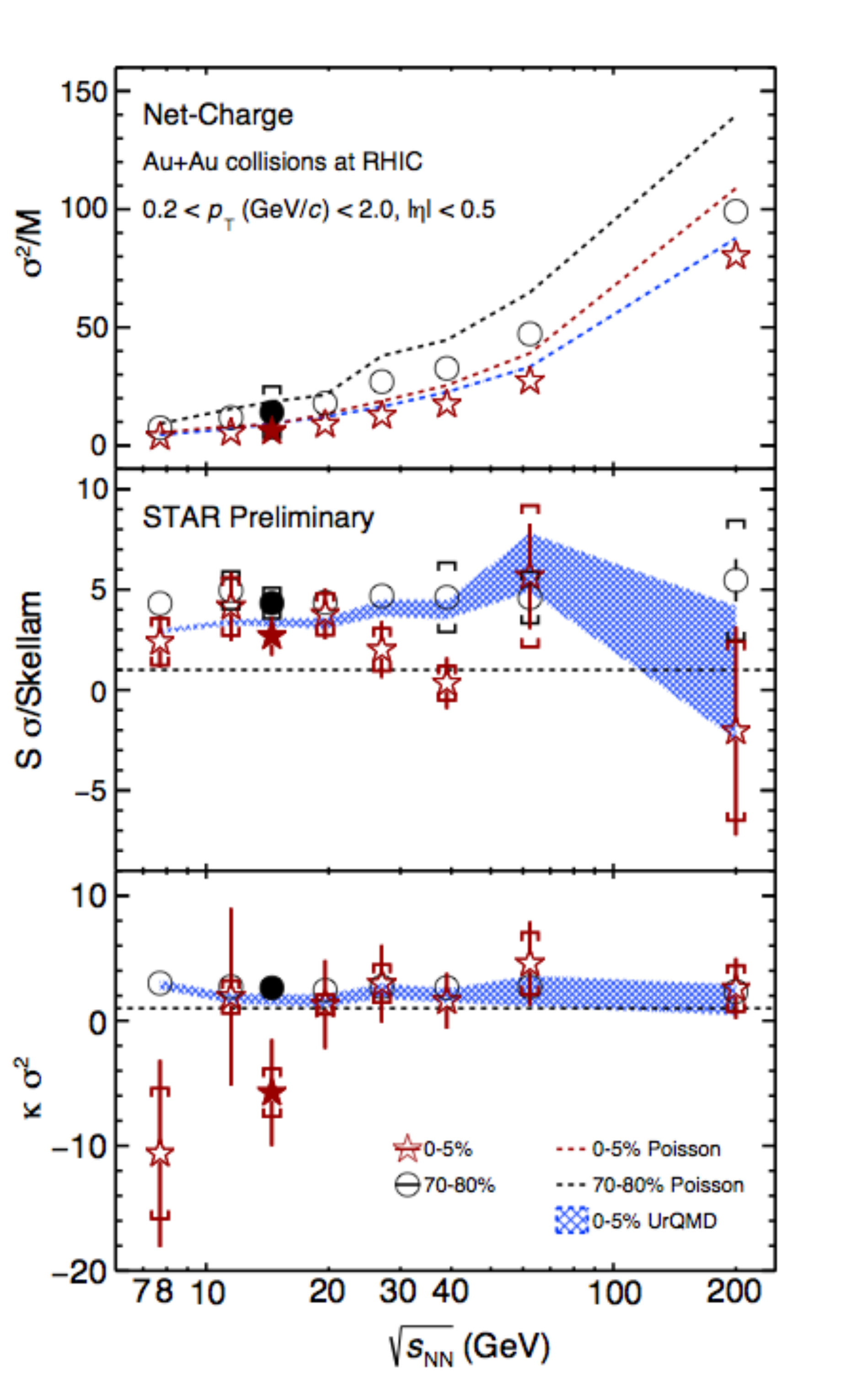}
\includegraphics[width=1.96in]{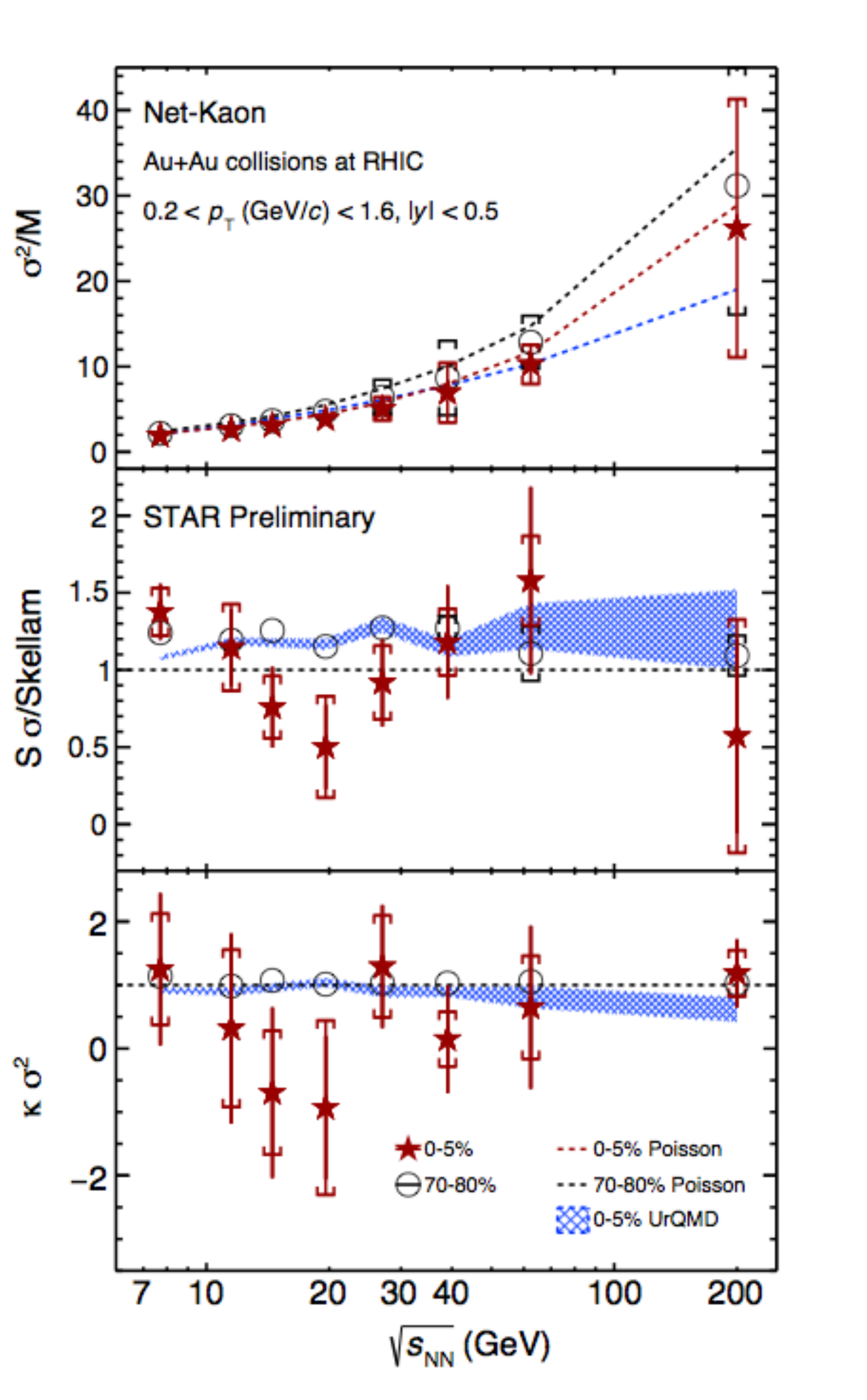}
\includegraphics[width=1.96in]{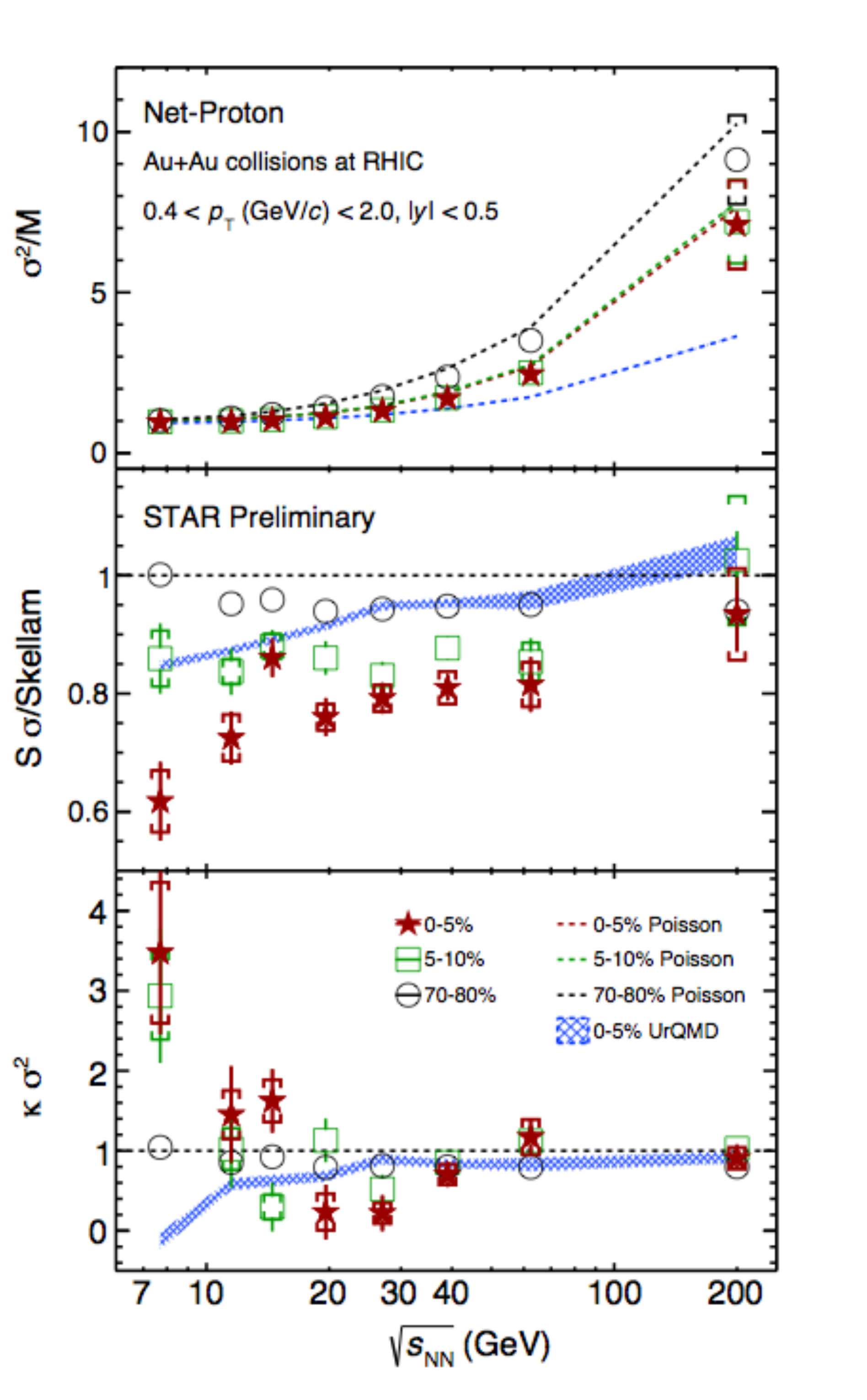}
\caption{Variation of the cumulant ratios for net-charge, net-kaon and net-proton multiplicity distribution with beam energy from the STAR experiment for different centralities.}
\label{fig:star}
\end{figure}

The various cumulant ratios can be written as
\begin{equation}
\frac{C_2}{C_1} = \frac{\sigma^2}{M}, \qquad \qquad \frac{C_3}{C_2} = S \sigma, \qquad \qquad \frac{C_4}{C_2} = \kappa \sigma,
\end{equation}
where $M$ is the mean, $\sigma^2$ is the variance, $S$ is the skewness and $\kappa$ is the kurtosis of the distribution.

Figure~\ref{fig:star} shows the measurements of the cumulant-ratios for net-charge (left panel), net-kaon (middle panel) and net-proton (right panel) multiplicity distributions from Au+Au collisions at $\sqrt{s_{NN}}$ = 7.7, 11.5, 14.5, 19.6, 27, 39, 62.4 and 200 GeV from the STAR experiment~\cite{fig1}. The black circles are the measurements for 70-80\%, green squares for 5-10\% and red stars for 0-5\% central collisions. The corresponding dashed lines are the Poisson baselines. The blue band is the result from 0-5\% central collisions from the UrQMD model calculations. It should be noted that the kinematic range for the measurements are different. Results for net-charge include all charged particles within transverse momentum region of 0.2 to 2.0 GeV/c and pseudorapidity ranging from -0.5 to 0.5. The spallation protons with transverse momentum less than 0.4 GeV/c are excluded. For the cumulants of net-kaons, (anti-)kaons within transverse momentum region of 0.2 to 1.2 GeV/c and rapidity ranging from -0.5 to 0.5 are included. For the net-proton analysis, (anti-)protons within transverse momentum region of 0.4 to 2.0 GeV/c and rapidity ranging from -0.5 to 0.5 are included. Within large statistical error bars, we find a monotonic trend in the various cumulant ratios for net-charge and net-kaon multiplicity distributions with the collision energy. However, higher cumulants of net-proton multiplicity distributions show a non-monotonic trend for central collisions. 

\begin{figure}[htb]
\centering
\includegraphics[width=4in]{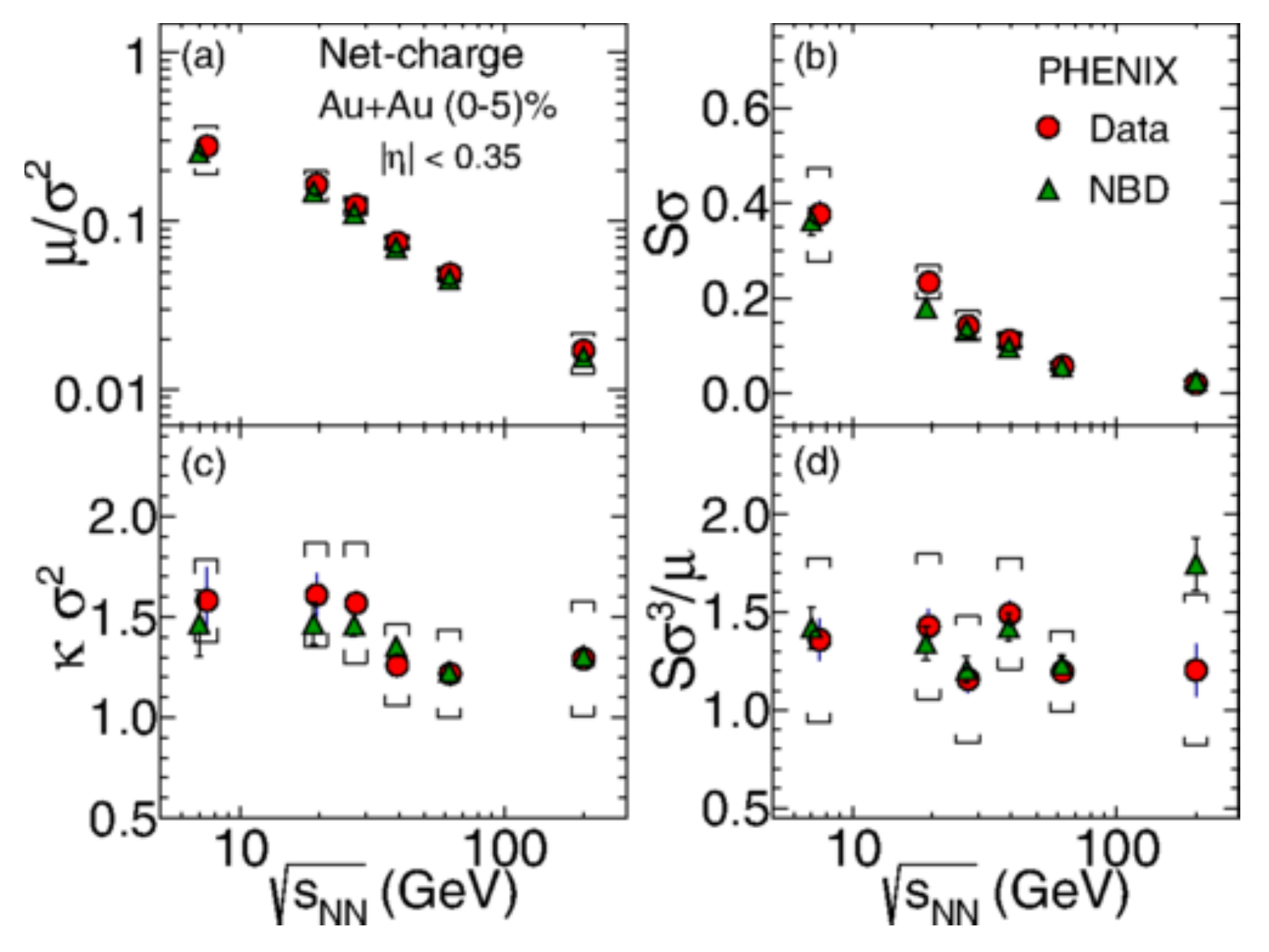}
\caption{Variation of the cumulant ratios for net-charge multiplicity distribution with beam energy from the PHENIX experiment for 0-5\% central collisions.}
\label{fig:phenix}
\end{figure}

Figure~\ref{fig:phenix} shows the various cumulant ratios of the net-charge multiplicity distribution for 0-5\% central collisions~\cite{fig2}. Within errors, the results are consistent with the expectations from Negative Binomial baseline. However, more statistics is needed at low beam energies.

\begin{table}[h]
\begin{center}
\caption{Relation between cumulants, ($C$) and correlation functions, ($\hat{\kappa}$)}
\begin{tabular}{lcl}
\label{tab:corr}
 $\hat{\kappa}_1  = C_1$ & $ \qquad $ & $C_1 = \langle N \rangle$  \\
 $\hat{\kappa}_2  = C_2 - C_1$ & $ \qquad $& $C_2 = \langle N \rangle + \hat{\kappa}_2 $  \\
 $\hat{\kappa}_3  = C_3 - 3C_2 + 2C_1$ & $  \qquad $ &$ C_3 = \langle N \rangle + 3 \hat{\kappa}_2 + \hat{\kappa}_3 $  \\
 $\hat{\kappa}_4  = C_4 - 6C_3 + 11C_2 - 6C_1$ & $  \qquad $ &  $C_4 = \langle N \rangle + 7 \hat{\kappa}_2 + 6 \hat{\kappa}_3 + \hat{\kappa}_4 $ 
\end{tabular}
\end{center}
\end{table}

\begin{figure}[htb]
\begin{center}
\includegraphics[scale=0.7]{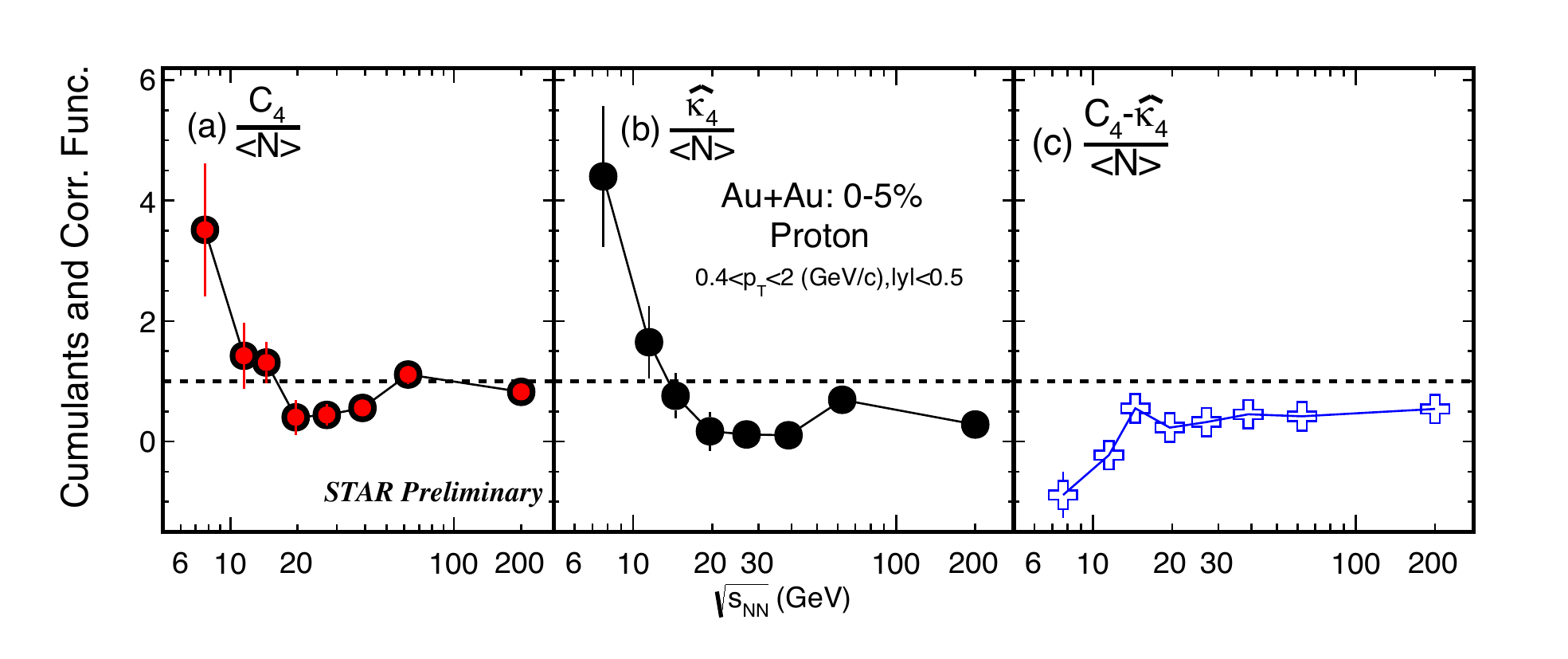}
\caption{Energy dependence of the cumulants and correlation functions of proton multiplicity distribution in central Au$+$Au collisions. The systematic uncertainties have not yet been studied.}
\label{fig:corr}
\end{center}
\end{figure}

In order to extract the multi-particle correlation functions ($\hat{\kappa}$) from the cumulants, we use the equations in Table~\ref{tab:corr}. Here, $\langle N \rangle$ is the average number of particles over the ensemble, and the $n^{th}$-order cumulants are expressed as a combination of lower order correlation functions ($\hat{\kappa}_n$). The correlation functions follow the same power-law dependence on correlation lengths as the cumulants. Figure~\ref{fig:corr} shows the normalized fourth-order cumulant for protons for 0-5\% central collisions with beam energy in the first panel and the normalized four-particle correlation function in the second panel~\cite{xfl}. Both of these show non-monotonic energy dependence. Upon subtracting $\hat{\kappa}_4$ from $C_4$, we find a monotonic energy dependence, which indicated that the non-monotonic behavior almost entirely arises from multi-particle correlations.

\begin{figure}[htb]
\begin{center}
\includegraphics[height=2in]{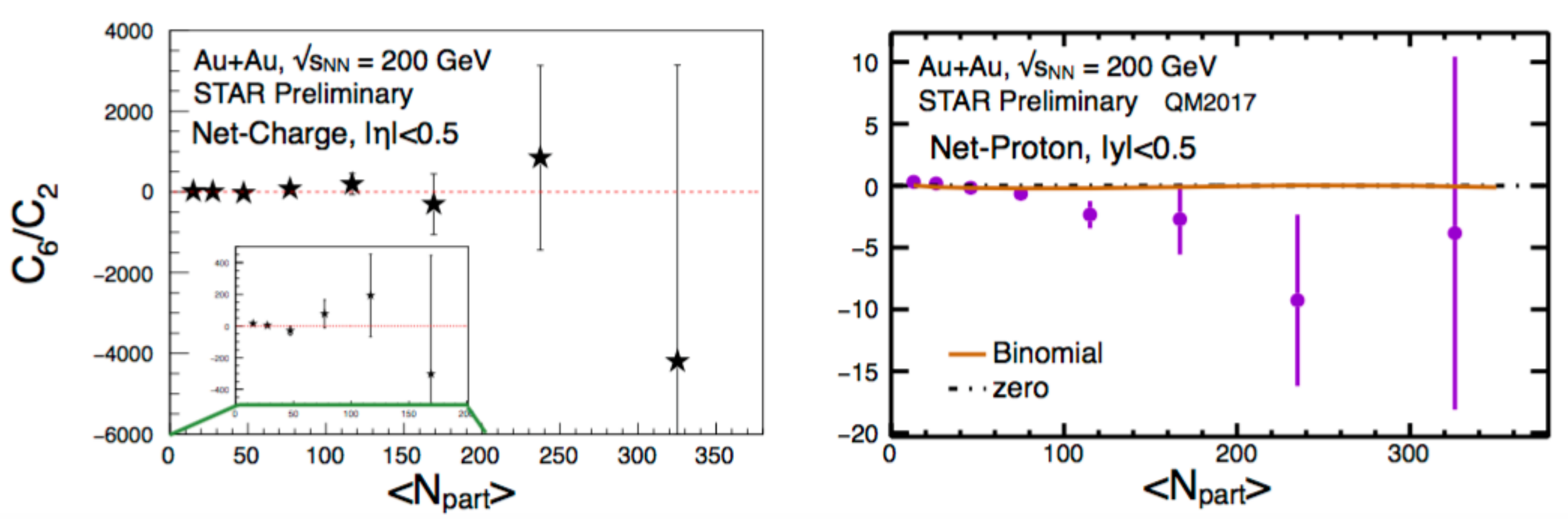}
\caption{Centrality dependence of the ratio of sixth-order to second-order cumulants of net-charge (left) and net-proton (right) multiplicity distribution in Au$+$Au collisions at $\sqrt{s_{NN}}$ = 200 GeV at STAR.}
\label{fig:c6}
\end{center}
\end{figure}

The sixth-order cumulants of net-charge and net-baryon distributions are predicted to be negative if the chemical freeze-out is close enough to the phase transition~\cite{lat2}. Figure~\ref{fig:c6} shows the measurement of the ratio of the sixth-order cumulant to the second-order cumulant for net-charge (left) and net-proton (right) multiplicity distributions for Au+Au collisions at $\sqrt{s_{NN}} = 200$ GeV at STAR~\cite{fig4}. We find that the ratio for net-charge is consistent with zero within large statistical errors. However, despite large error bars, negative values are observed for $C_6/C_2$ for net-proton systematically from peripheral to central collisions. The systematic uncertainties have not yet been determined.

\section{Unfolding for efficiency correction}
Experimental effects like particle misidentification, track merging/splitting, etc. could lead to non-Binomial detector efficiencies. In addition, previous studies have shown that there could be noticeable consequences of multiplicity-dependent behavior of detection efficiency on higher-order cumulants~\cite{mc1}. In order to understand these, unfolding techniques are being developed at STAR for reliable measurement of the cumulants. 

The measured distribution is a convolution of original particle distribution and the detector response function. There are two key elements of the approach, namely, the correlation matrix and the response histogram. The correlation matrix contains the number correlation between the measured protons and anti-protons. The response histogram contains the distribution of produced particles for every detected number of particles. These are obtained using embedding. Two schemes for unfolding are being looked into - unfolding with initial (anti-)proton distributions assumed to be Poisson distributions~\cite{met1} and unfolding with iterations~\cite{met2}.

\begin{table}[h]
\begin{center}
\caption{Comparison of cumulants of net-proton multiplicity distribution using AMPT model with multiplicity-dependent efficiency for 0 -- 5\% central Au+Au collisions at $\sqrt{s_{NN}}$ = 200 GeV}
\begin{small}
\begin{tabular}{|c|c|c|c|c|}
\hline 
Cumulants for  & True  & Efficiency corrected & Efficiency corrected & Efficiency corrected \\
net-proton &  distribution & (2-D response & (1-D response & (Factorial moment \\
distribution & & matrix) & matrix) & method) \\ 
\hline 
$C_1$ & 2.799 $\pm$ 0.002 & 2.799 $\pm$ 0.002 & 2.800 $\pm$ 0.002 & 2.550 $\pm$ 0.001\\
$C_2$ & 31.44 $\pm$ 0.01 & 31.43 $\pm$ 0.01 & 49.78 $\pm$ 0.02 & 12.63 $\pm$ 0.01 \\
$C_3$ & 8.4 $\pm$ 0.2 & 8.4 $\pm$ 0.1 & 9.3 $\pm$ 0.2 & 2.58 $\pm$ 0.04\\
$C_4$ & 91 $\pm$ 1 & 91 $\pm$ 2 & 89 $\pm$ 3 & 12.5 $\pm$ 0.3 \\ \hline
\end{tabular}
\end{small}
\label{tab:comp}
\end{center}
\end{table}

As an illustration, we use the event-by-event distribution of (anti-)protons given by the AMPT model. The efficiency for protons is assumed to be 0.8 -- 0.0003 $(N_{charge} - N_{proton} - N_{anti-proton})$, while for anti-protons, it is 0.7 -- 0.0003 $(N_{charge} - N_{proton} - N_{anti-proton})$ respectively. The coefficient 0.0003 is the expected order of magnitude of the multiplicity dependence of efficiency in real data. We compare the results in Table~\ref{tab:comp}. We find that the results from the data-driven method agree well with the cumulants of the true distribution. In the 2-D response matrix situation, both protons and anti-protons are corrected simultaneously, that is, the response matrix is a two-dimensional histogram containing the information for produced number of both protons and anti-protons for every measured number of protons and anti-protons. In the 1-D response matrix approach, protons and anti-protons are corrected separately, that is, the response matrix is two one-dimensional matrices; one for protons and the other for anti-protons. The corrected cumulants obtained from the factorial moment method, however, deviate from the true values considerably. This is because the factorial moment method assumes binomial efficiency correction. Centrality bin width correction is applied to the factorial moment method. Details of the efficiency correction appear to have a significant effect on the cumulants and STAR is in the process to determine a suitable analysis approach.

\section{Summary and outlook}
The cumulant-ratios for net-charge, net-kaon and net-proton multiplicity distributions were presented from both the STAR and PHENIX experiment from the Beam Energy Scan program at RHIC. We found that the ratios were mostly monotonic within error bars. Non-monotonic energy dependences are observed for $C_4/C_2$ of protons and net-proton multiplicity distributions for 0-5\% central Au+Au collisions. We found that the four-particle correlations contribute dominantly to the observed non-monotonicity. The ratio $C_6/C_2$ is consistent with being negative as expected from theoretical calculations for Au+Au collisions at $\sqrt{s_{NN}} = 200$ GeV within large statistical uncertainties.

Efficiency correction is an important ingredient in order to reliably calculate the higher-order cumulants. We are currently developing unfolding methods to account for various effects in detection efficiencies. Such an approach would be a significant improvement over the previous analyses under the assumption of binomial efficiency. AMPT simulations indicated that there could be a considerable effect on the cumulant calculations. Unfolding methods are being developed at STAR to account for these.

More data will be collected in BES-II for Au+Au collisions at $\sqrt{s_{NN}}$ = 7.7 -- 19.6 GeV in 2019 -- 2020 with detector upgrades.

\end{document}

%% file: econfmacros.tex
\def\beq{\begin{equation}}
\def\eeq#1{\label{#1}\end{equation}}
\def\eeqn{\end{equation}}

\def\beqa{\begin{eqnarray}}
\def\eeqa#1{\label{#1}\end{eqnarray}}
\def\eeqan{\end{eqnarray}}

\let\bar=\overbar

\def\Dslash{\not{\hbox{\kern-4pt $D$}}}
\def\dslash{\not{\hbox{\kern-2pt $\del$}}}

\def\msb{{\bar{\ssstyle M \kern -1pt S}}}

